\documentclass[manuscript]{aastex62}

\usepackage{natbib}
\usepackage{color}
\usepackage{multirow}
\shorttitle{Planetary Radius, Gravity, and Climate}
\shortauthors{Yang \& Yang}

\usepackage{hyperref}
\hypersetup{colorlinks=true,linkcolor=black,citecolor=blue,urlcolor=black}

\begin{document}

\title{How do Planetary Radius and Gravity Influence the Surface Climate of Earth-like Planets?}

\correspondingauthor{Jun Yang}
\email{junyang@pku.edu.cn}

\author{Huanzhou Yang$^{1}$, and Jun Yang} \affil{Dept. of Atmospheric and Oceanic Sciences, School of Physics, Peking University, Beijing, 100871, China\\
} 

\begin{abstract}
About 4000 exoplanets have been confirmed since the year of 1992, and for most of the planets, the main parameters that can be measured are planetary radius and mass. Based on these two parameters, surface gravity can be estimated. In this work, we investigate the effects of varying radius and gravity on the climate of rapidly rotating terrestrial planets with assuming they have an ocean and Earth-like atmospheres (N$_2$, CO$_2$, and H$_2$O). Using a three-dimensional (3D) atmospheric general circulation model (GCM), we find that varying radius mainly influences the equator-to-pole surface temperature difference while varying gravity mainly influences the mean surface temperature. For planets of larger radii, the meridional atmospheric energy transport is weaker, which warms the tropics but cools high latitudes. For planets of larger gravities, the surface is globally cooler due to the fact that saturated vapor pressure depends on air temperature only and column water vapor mass is approximately equal to saturated vapor pressure over gravity multiplied by relative humidity. The relative humidity does not change much in our experiments. Ice albedo and water vapor feedbacks act to further amplify the effects of varying radius and gravity. These results suggest that radius and gravity are important factors for planetary climate, although they are not as critical as stellar flux, atmospheric composition and mass, and rotation rate. Future simulations are required to know how large the radius and gravity could influence the width of the habitable zone using 3D GCMs. \\
\end{abstract}
\keywords{planets and satellites: general --- planets and satellites: atmospheres --- methods: numerical --- astrobiology}


\section{Introduction} \label{sec:Itr}
Finding potentially habitable planets beyond the solar system is the prime target of exoplanet missions. These planets should have sizes or masses similar to Earth and meanwhile be in the habitable zone within which planetary surface is temperate to maintain liquid water \citep{KASTING1993108}. Previous studies have shown that various factors could influence the surface climate of planets in the habitable zone, including the stellar flux on the planet, stellar spectrum, atmospheric composition and mass, planetary rotation rate, planetary obliquity, orbital eccentricity, etc.~\citep{Kopparapu_2013, Yang_2013, Turbet2016, Wolf_2017, Salameh2018, Genio2019}. However, the effects of planetary radius and gravity have rarely been studied although these two parameters much better known than other parameters in observations. With doppler and transit techniques, the mass (or the minimum mass) and the radius of exoplanets can be measured, although significant uncertainties exist \citep{2014ApJ...783L...6W}. For the radii and gravities of several potentially habitable planets, please see Figure~1. 



\begin{figure}[!htbp]
\begin{center}
\includegraphics[angle=0, width=8.5cm]{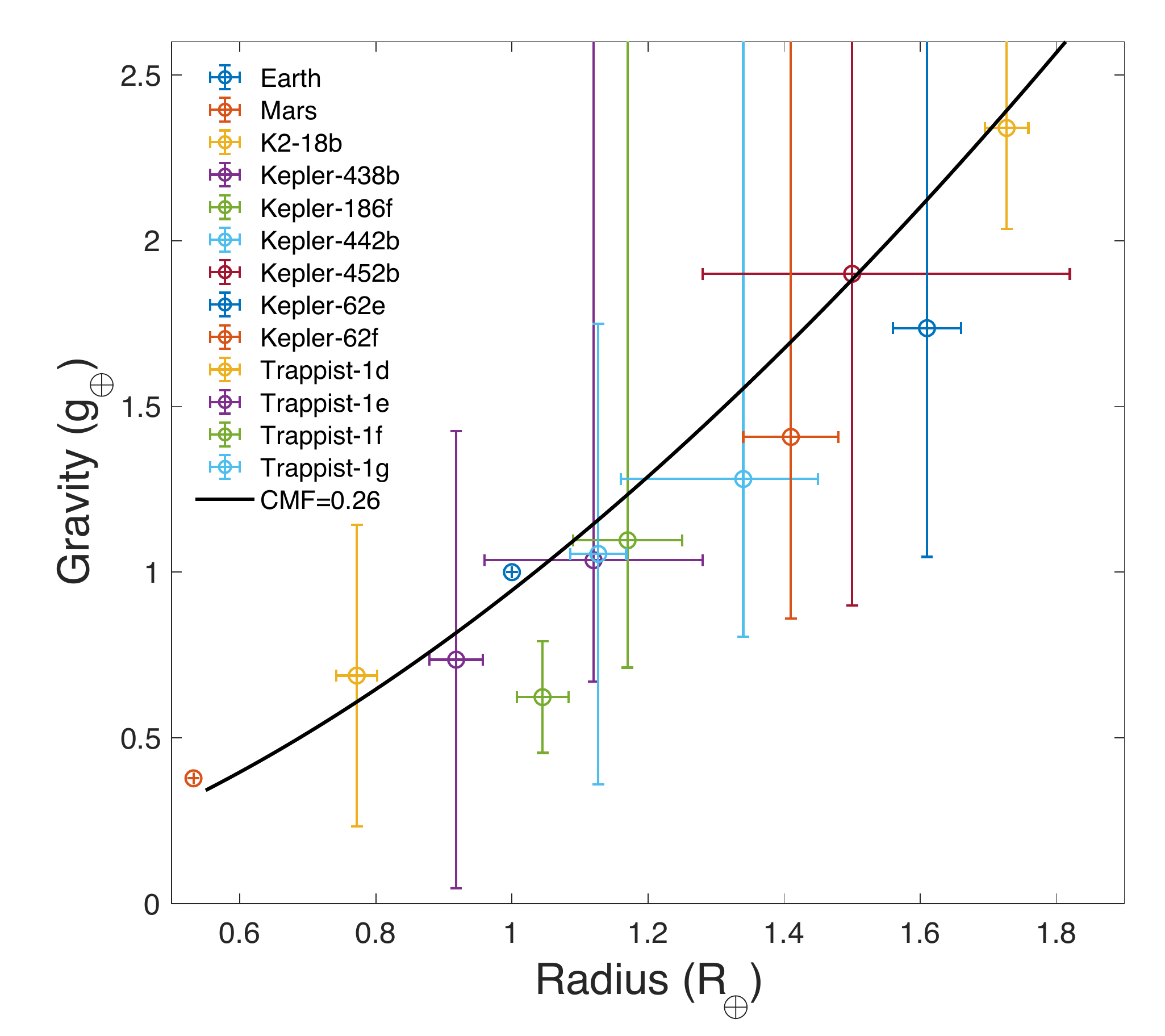}
\caption{Planetary radius and gravity distribution of several confirmed exoplanets. The values of Earth and Mars and the uncertainties are shown. The black line represents the theory relation with a core mass fraction of Earth's value, 0.26 \citep{Zeng_2016}.}
\label{RadiusGravity}
\end{center}
\end{figure}


\cite{Kopparapu_2014} explored the effect of planetary masses on the habitable zone using a 1D radiative transfer model. They found that the inner edge of the habitable zone for planets with larger masses are closer to the star because water vapor column path becomes smaller. Using an idealized 3D GCM with a gray radiation scheme and without the feedbacks of water vapor, cloud, or ice albedo, \cite{Kaspi_2015} investigated the effects of various parameters (rotation rate, stellar flux, atmospheric mass, atmospheric optical thickness, surface gravity, and radius) on the atmospheric circulation and the equator-to-pole surface temperature difference on terrestrial planets orbiting G-type stars. They found that larger planets have a smaller meridional surface temperature gradient (in units of K~km$^{-1}$) but a larger equator-to-pole temperature difference (in a unit of K). \cite{Kilic_2017b} investigated the effect of varying gravity on the atmospheric circulation using an idealized GCM with fixed surface temperatures. \cite{Komacek:2019aa} did a similar work to \cite{Kaspi_2015} but used a more complex GCM that includes the effects of cloud, water vapor feedback, nongray radiative transfer, and sea ice; moreover, they simulated planets orbiting both Sun-like  and M-type stars. They found that the equator-to-pole surface temperature difference is enhanced for larger planets orbiting Sun-like stars; this trend is the same as that found in \cite{Kaspi_2015}. They also found that the equator-to-pole temperature difference increases with increasing gravity but the underlying mechanism was not addressed. 


In this work we analyze how radius and gravity influence the surface climate of rapidly rotating planets using a 3D GCM with nongray radiation transfer. Compared to the idealized model used in \cite{Kaspi_2015}, the model we employed is better in investigating the response of the climate because water vapor, cloud, and ice albedo feedbacks could be included in our simulations. Compared to the work of \cite{Komacek:2019aa}, we will more clearly show the underlying mechanisms for the effects of varying radius and gravity on the surface climate. 

The outline of this paper is as follows. In section \ref{sec:Met}, we describe the model and experimental designs. In section \ref{sec:Res}, we demonstrate the results and analyze the underlying mechanisms. In most of our analyses, surface temperature is the key variable, and atmospheric circulation is briefly addressed in case to understand the response of the meridional surface temperature gradient. In section \ref{sec:Dis}, we discuss the effect of cloud feedback and the possible effect of varying radius and gravity on wind-driven ocean circulation. Conclusions are stated in section \ref{sec:Con}. \\



\section{Model descriptions and Experimental Designs} \label{sec:Met}
We employ the 3D GCM---the Community Atmospheric Model version 3, {\tt CAM3}. The model was developed to simulate the climate of Earth in the present, past, and future  \citep{Collins:2004aa}. We have modified the model to simulate the climates of various planets in the habitable zone around different stars \citep{Yang_2013,Yang:2017aa}. The model solves the primitive equations for atmospheric motion on a rotating sphere and the equations for radiative transfer, convection, condensation, precipitation, cloud, and boundary turbulence \citep{Bovile:2006}. The model is coupled to a 50 m mixed layer ocean module and a thermodynamic sea ice module. The surface is set to be seawater everywhere, i.e., an aquaplanet. Meridional oceanic heat transport is specified to be similar to modern Earth but it is set to be symmetry about  the equator. Neither ocean nor sea ice dynamics are considered.

 
We perform three groups of experiments. In the first and second groups, we vary the radius and gravity separately, in order to isolate the dependence of the climate on each parameter. In the third, we modify the radius and gravity simultaneously. Three values of radius ($R$) were examined, 0.5, 1, and 2 of Earth's value (6371 km), denoted as 0.5$R_\oplus$, 1$R_\oplus$, and 2$R_\oplus$, respectively. Three values of gravity ($g$) have also been investigated, 0.5$g_\oplus$, 1$g_\oplus$, and 2$g_\oplus$ ($g_\oplus$\,=\,$9.8\,\,$m$\,$s$^{-1}$). Results comparing 0.5$R_\oplus$ and 1$R_\oplus$ and comparing 1$R_\oplus$ and 2$R_\oplus$ are similar in surface temperature changes, so that for concision we show the results of the latter, and the same for gravity. 

Two types of experiments were performed, one with neither sea ice nor ice albedo feedback (the freezing point is artificially set to an extremely low value, such as 100~K), and the other one with sea ice and ice albedo feedback (the freezing point is 271.35~K, or -1.8$^{\circ}$C). The first type is set in order to more clearly know the direct effects of varying radius and gravity, and the second type is set in order to know the effect of ice/snow albedo feedback. Sea ice albedo is 0.68 in the visible band and 0.30 in the near-infrared band, and snow albedo is 0.91 and 0.63, respectively \citep{briegleb2002}.


Atmospheric compositions are set to be Earth-like, a N$_2$-dominated atmosphere with CO$_2$ and H$_2$O but no ozone. CO$_2$ mixing ratio is 600 ppmv (this value does not influence our conclusion). When the gravity is modified, we assume the column air mass (as well as column CO$_2$ mass) does not change, so that the surface pressure varies. For instance, the surface pressure is 1.0 and 2.0 bar
for the gravity of 1$g_\oplus$ and 2$g_\oplus$, respectively, but the column air mass ($\approx$10$^4$~kg\,m$^{-2}$) and the column CO$_2$ mass ($\approx$9.4~kg\,m$^{-2}$) are the same between the experiments. An additional experiment with a 1 bar atmosphere and 1200 ppmv CO$_2$ under 2$g_\oplus$ is also performed and we find that it does not significantly influence our results (see Figure~3). Each run is initialized from a state close to modern Earth but without sea ice. Each experiment runs for about 60 Earth years and the final 5 years are used for the analyses shown hereafter.

Orbital parameters are set as simplified as possible. Both obliquity and eccentricity of the orbit are set to zero. Rotation period and orbital period are equal to those of Earth. The solar radiation is 1365 W\,m$^{-2}$ and the spectrum of the Sun is used. So that, the results shown here are for rapidly rotating planets around G-type stars. In a separate paper, we address the effect of varying radius and gravity on the climates for tidally locked planets around M stars \citep{Yang_2019}.\\


\section{Results} \label{sec:Res}

\subsection{Effects of Varying Planetary Radius} \label{sec:Radius}

\begin{figure}[!htbp]
\begin{center}
\includegraphics[angle=0, width=20cm]{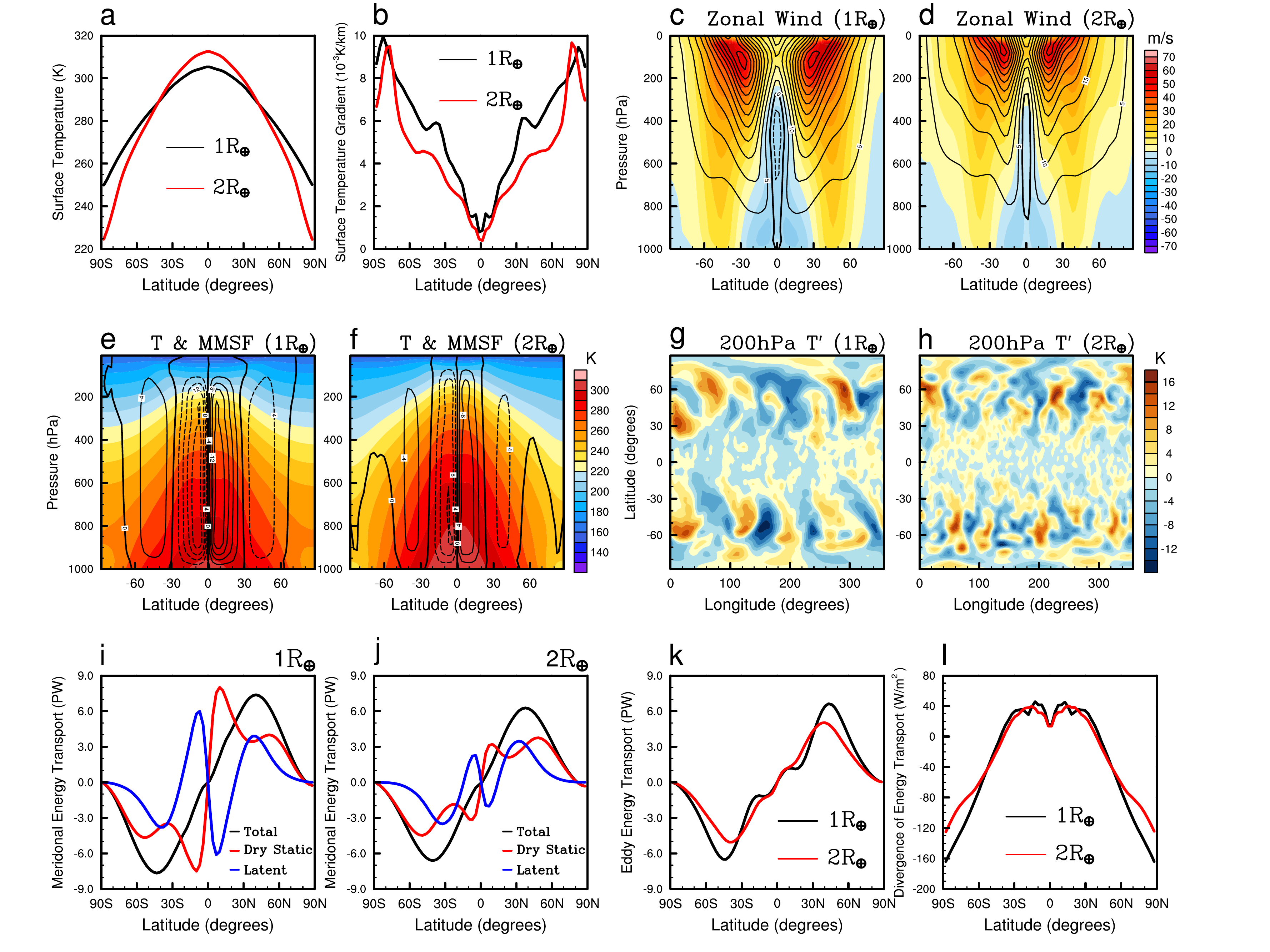}
\caption{Effects of varying radius on the climate of Earth-like aqua-planets. (a) Zonal-mean surface temperatures. (b) Meridional gradient of the surface temperatures in units of 10$^{-3}$~K~km$^{-1}$. (c \& d) Zonal-mean zonal winds (color shading) and the corresponding zonal winds calculated based on thermal wind balance (contour lines). (e \& f) Zonal-mean air temperature (color shading) and zonal-mean mass streamfunction in units of 10$^{11}$ kg\,s$^{-1}$ (contour lines). (g \& h) Transient air temperature anomaly at the level of 200\,hPa (the zonal-mean values are subtracted). (i \& j) Zonal-mean meridional transport (PW, 1~PW = 10$^{15}$ W) of dry static energy, latent heat, and total energy. (k) Zonal-mean meridional energy transport by transient eddies. Stationary eddies can be ignored because of the absence of continents and seasonal cycles in our experiments. (l) Zonal-mean net radiation flux (shortwave minus longwave) at the top of the atmosphere. In order to more direct comparisons, the values for the $2R_\oplus$ case in (i--k) are divided by a factor of 4, due to the fact that its sphere area is four times that in the $1R_\oplus$ case. Note that the ice and ice albedo feedback are turned off in these experiments.}
\label{RadiusEffect}
\end{center}
\end{figure}



When increasing the planetary radius, the main characterization of the surface temperature changed is  that the equator becomes warmer and the polar regions become cooler (Figure~2(a)), which means that the temperature difference between the equator and the poles in the unit of K becomes larger. However, the surface temperature gradient ($\frac{\partial{T}}{\partial{y}}$ = $\frac{\partial{T}}{R\partial{\phi}}$, where $R$ is the planetary radius and $\phi$ is the latitude) in the unit of K~km$^{-1}$ decreases (Figure~2(b)), which is primarily due to the geometrical effect of increasing $R$. Note that in discussing the temperature contrast for different radii, the units are important. The change of global-mean surface temperature between these two cases is small, only 1.1~K. Due to the decrease of the surface temperature gradient, all of the atmospheric jets, tropical Hadley cells, and mid-latitude Ferrel cells and transient eddies become weaker (Figure~2(c--h)). As a result, the meridional energy transport, including both dry static energy and latent heat, decreases (Figure~2(i--k)). This means less energy is transported from the tropics to the polar regions, so that the surface temperatures in the deep tropics increase and meanwhile the surface temperatures in high latitudes decrease. The area-mean decrease in the divergence of the meridional energy transport is 5.8 W~m$^{-2}$ (Figure~2(l)).

Comparing Figure 2(e) to Figure 2(f), one could find that the Hadley cells become narrower in the unit of degrees latitude ($^\circ$) but become wider in the unit of distance (m) when the radius is increased from 1$R_\oplus$ to 2$R_\oplus$. The change of the Hadley Cell width can be understood using the theory of \cite{PHILLIPS:1954aa} and \cite{Held2001}:  
\begin{equation}
\phi_h \propto (\frac{NH}{\Omega R})^\frac{1}{2},~~or~~D_h \propto (\frac{NHR}{\Omega})^\frac{1}{2},
\end{equation}
where  $N$ is the buoyancy frequency, $H$ is the scale height, and $\Omega$ is the rotating rate. Clearly, the value of $\phi_h$ decreases whereas the value of $D_h$ increases with increasing $R$. While the width of the Hadley cells is connected to $R$, the characteristic length scale of mid-latitude eddies is not. It is determined by nature properties of the atmosphere. For example, the first deformation radius of baroclinic eddies is equal to $NH/f$, where $f$ is the Coriolis frequency \citep{Vallis06}. When the planetary radius is changed, the values of $N$ and $H$ exhibit very small changes, so that the length scale of mid-latitude eddies in the unit of meters nearly does not change (Figure 2(g--h)). The eddy length scale relative to the value of $R$, however, becomes smaller, which also promotes a weaker meridional energy transport in mid-latitudes, as previously pointed out in \cite{Kaspi_2015}. \\


\subsection{Effects of Varying Planetary Gravity}  \label{sec:Gravity}

When the gravity is increased, the surface temperatures decrease, regardless of whether  the column air mass or the surface pressure is fixed (Figure~3(a)). In global mean, the surface temperature decreases by 8 K comparing ($2g_\oplus$, 2bar) and ($1g_\oplus$, 1bar). The main reason for the surface cooling is the decreasing of water vapor content (Figure~3(b)) and thereby the clear-sky greenhouse effect falls (Figure~3(c)). For a pure water vapor atmosphere, the water vapor mass per unit area is inversely proportional to gravity: $e_s/g$, where $e_s$ is the saturation vapor pressure and it only depends on air temperature. For an atmosphere with background gas (such as N$_2$), the column water vapor mass per unit area is: 
\begin{equation}
{M}=\int_{0}^{P_0} q\frac{dp}{g} \approx \,\,   \frac{\varepsilon}{g}  \,  \int_{0}^{P_0}  \gamma e_s \frac{dp}{p},
\end{equation}
where $q$ is the specific humidity, $P_0$ is the surface pressure, $\gamma$ is the relative humidity, $\varepsilon$ is the ratio of H$_2$O molecular mass to total air molecular mass ($\varepsilon = 0.622$ for the present-day Earth), $p$ is the air pressure, and $dp$ is the pressure change between two levels. From this equation, it is clear to see that the column vapor mass is approximately inversely proportional to gravity, as long as $\gamma$ has no significant change. Indeed, $\gamma$ does not change much in the experiments. Therefore, less (more) water vapor is required to maintain a given vapor pressure for a larger (smaller) gravity, as discussed in \cite{Pierrehumbert:2010} and \cite{Kopparapu_2014}.

\begin{figure}[!htbp]
\begin{center}
\includegraphics[angle=0, width=18.5cm]{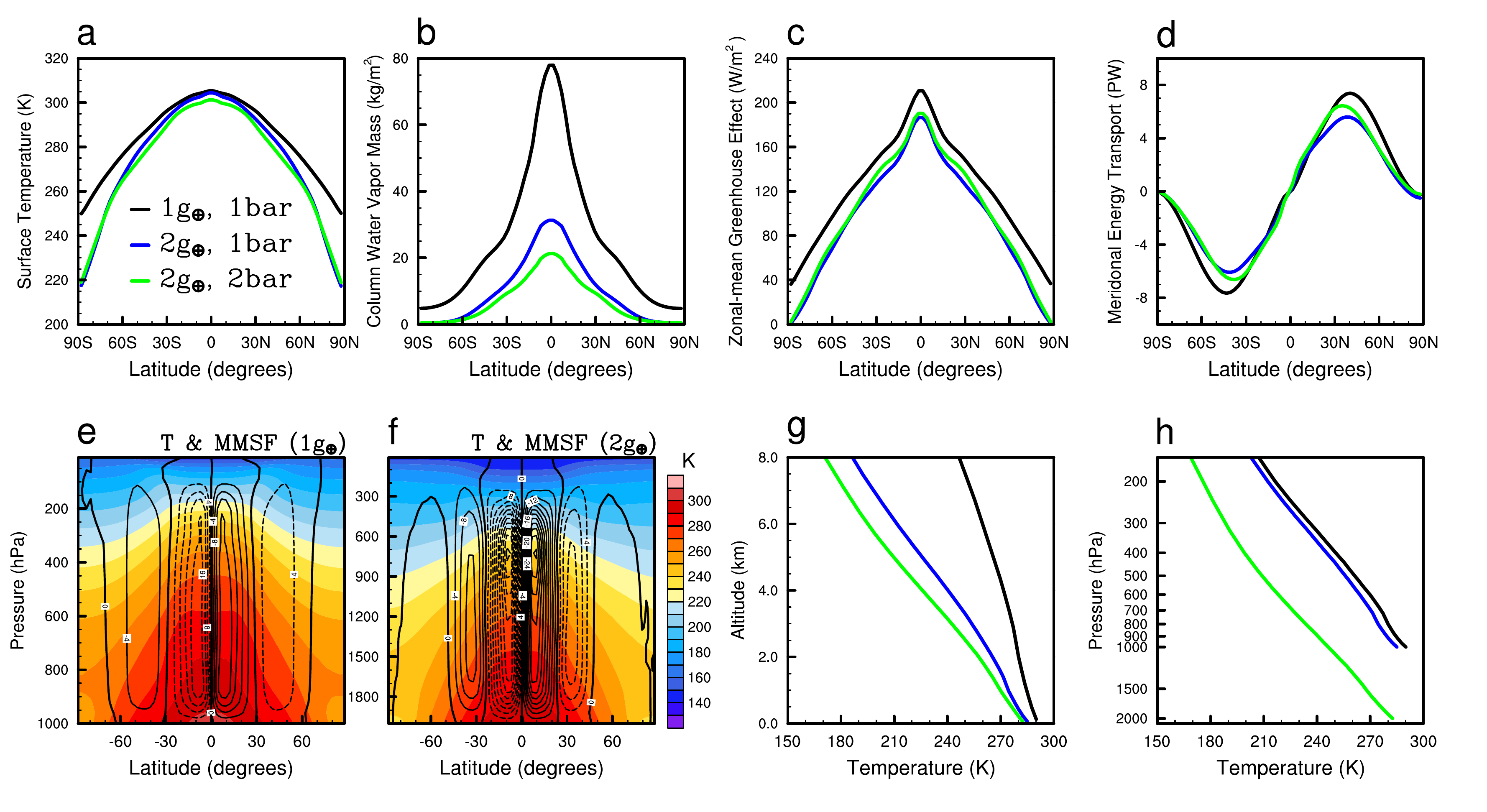}
\caption{Effects of varying gravity on the climate of Earth-like aqua-planets. (a) Zonal-mean surface temperatures. (b) Vertically-integrated water vapor mass. (c) Atmospheric clear-sky greenhouse effect. (d) Meridional total energy transport. (e \& f) Air temperature (color shading) and mass streamfunction in units of 10$^{11}$ kg\,s$^{-1}$ (contour lines). (g \& h) Global-mean air temperature profile in the z-coordinate and in the p-coordinate. Note that the ice and ice albedo feedback are turned off in these experiments.}
\label{GravityEffect}
\end{center}
\end{figure}



Beside the decrease in the global-mean surface temperature, the cooling in high latitudes is greater than that in the tropics (a `polar amplification'), i.e., the meridional temperature gradient increases. The underlying mechanisms are essentially the inverse of a global-warming response, including the water vapor feedback, cloud feedback, lapse rate feedback, Planck feedback, response of meridional energy transport, and changes of surface latent and sensible heat fluxes, as illustrated for instance in \cite{PithanandMauritsen2014}. The dominating mechanism is that the meridional energy transport diminishes significantly in mid-latitudes when increasing gravity (Figure~3(d)), promoting a greater cooling in high latitudes. The reducing of the energy transport mainly results from the decrease of latent heat transport (figure not shown). The Hadley and Ferrel cells, however, become stronger, the same as that found in the studies of \cite{Kaspi_2015} and \cite{Komacek:2019aa}; this is due to the increasing of the temperature gradient. 

Global-mean temperature profiles for different gravities are shown in Figure 3(g-h). In the $z$ coordinate, the lapse rate ($\Gamma = - \frac{dT}{dz}$) increases with gravity. This is due to the change in the scale height, which is inversely proportional to gravity. In the $p$ coordinate, however, the lapse rate ($\frac{dT}{dp}$) nearly does not change. 

Comparing the cases ($2g_\oplus$, 2bar) and ($2g_\oplus$, 1bar), one could find that the surface temperature decreases in the tropics but increases slightly in the polar regions as increasing the air mass. This is due to the combined effect of enhanced atmospheric Rayleigh scattering and increased meridional energy transport from the tropics to the polar regions (Figure~3(d)). The clear-sky planetary albedos are 0.17 and 0.13 in the cases of ($2g_\oplus$, 2bar) and ($2g_\oplus$, 1bar), respectively, and  for given atmospheric velocities, the meridional energy transport could be higher for a larger air mass. 


When we were writing this manuscript, \cite{Thomson:2019aa} published a gravity--climate paper that has the same conclusion shown here. Using theoretical arguments and 3D GCM simulations, they found  that the same mechanism: a larger gravity leads to a less condensible (such as water vapor) and thereby a weaker greenhouse effect, so that the surface becomes cooler.  \\



\subsection{Results Including Ice Albedo Feedback}

When sea ice is included, the climatic response to varying radius and gravity becomes greater (Figure~4). When only the radius is doubled, ($2R_\oplus$, $1g_\oplus$) versus ($1R_\oplus$, $1g_\oplus$), the surface temperature increases in the deep tropics but decreases in the mid- and high-latitudes. This is primarily due to the weakening of the meridional energy transport (Figure~4(d)), as discussed in the section~3.1. The cooling in the mid- and high-latitudes is larger than that in the simulations without ice albedo feedback (see Figure~2(a)). This is because that the sea ice edges expand toward the equator (Figure~4(b)), increasing the surface albedo and thereby cooling the system. Water vapor feedback acts to further cool the mid- and high-latitudes where the water vapor amount decreases but warms the deep tropics where it increases (Figure~4(c)). In global mean, the surface temperature decreases by 14 K. 

\begin{figure}[!htbp]
\begin{center}
\includegraphics[angle=0, width=18.3cm]{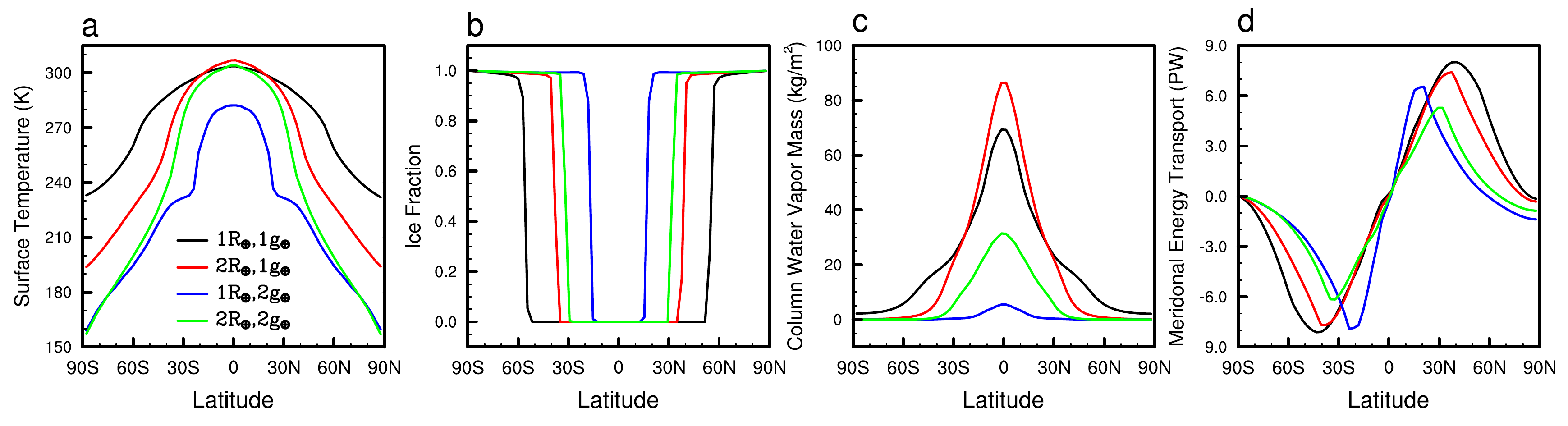}
\caption{Effects of varying radius and gravity on the climate of Earth-like aqua-planets with ice albedo feedback considered. (a) Zonal-mean surface temperature, (b) zonal-mean sea ice  fraction, (c) vertically integrated water vapor mass, and (d) meridional total energy transport. In (d), the values of ($2R_\oplus$, $1g_\oplus$) and ($2R_\oplus$, $2g_\oplus$) are divided by a factor of 4, same as that in Figure~2.}
\label{IceAlb}
\end{center}
\end{figure}



When only the gravity is doubled, ($1R_\oplus$, $2g_\oplus$) versus ($1R_\oplus$, $1g_\oplus$), the surface temperature decreases everywhere (blue line in Figure~4), and the global-mean surface temperature decreases by 50 K. The dominated mechanism is the decrease of water vapor amount with gravity, as discussed in the section~3.2. Ice albedo feedback acts to further cool the surface. The meridional energy transport decreases and its peak shifts toward the equator. This is due to that less solar radiation is absorbed by water vapor in the air and by the surface, so that less energy is required to be transported poleward. 

When the radius and gravity are doubled simultaneously, the surface temperature decreases everywhere, ($2R_\oplus$, $2g_\oplus$) versus ($1R_\oplus$, $1g_\oplus$), except in the deep tropics where the surface temperature nearly does no change (green line in Figure~4). In global mean, the surface temperature decreases by 28 K. The negligible change of surface temperature in the deep tropics is due to the competition between the cooling effect of decreasing water vapor amount and the warming effect of reducing meridional energy transport. \\


\section{Discussion} \label{sec:Dis}

Cloud is an important component of the climate system but it has a high uncertainty in GCM simulations \citep[e.g.,][]{Yangetal2019}. Figure~5(a) shows the net cloud radiative effect in the experiments with ice turned off. In global mean, its changes are within 6 W~m$^{-2}$, much smaller than the changes of clear-sky greenhouse effect, 20--40 W~m$^{-2}$ (Figure~3(c)). As increasing gravity, cloud water path decreases significantly especially in the deep tropics (figure not shown) because less water vapor can be maintained in the atmosphere as addressed in section~3.2.

Ocean circulation could also be influenced by planetary radius and gravity. Figure 5(b) shows the zonal surface wind stresses. When the radius is increased, the wind stresses generally become smaller, due to the decrease of surface temperature gradients as addressed in section~3.1. This implies that the wind-driven ocean circulation would become weaker. When the gravity is increased, the wind stresses generally becomes greater, due to the increase of surface temperature gradients as addressed in section~3.2. This implies that the wind-driven ocean circulation would become stronger. Future work is required to quantitatively understand the net effect of varying radius and gravity on the ocean circulation and meridional ocean heat transport using fully coupled atmosphere--ocean GCMs.\\

\begin{figure}[!htbp]
\begin{center}
\includegraphics[angle=0, width=14cm]{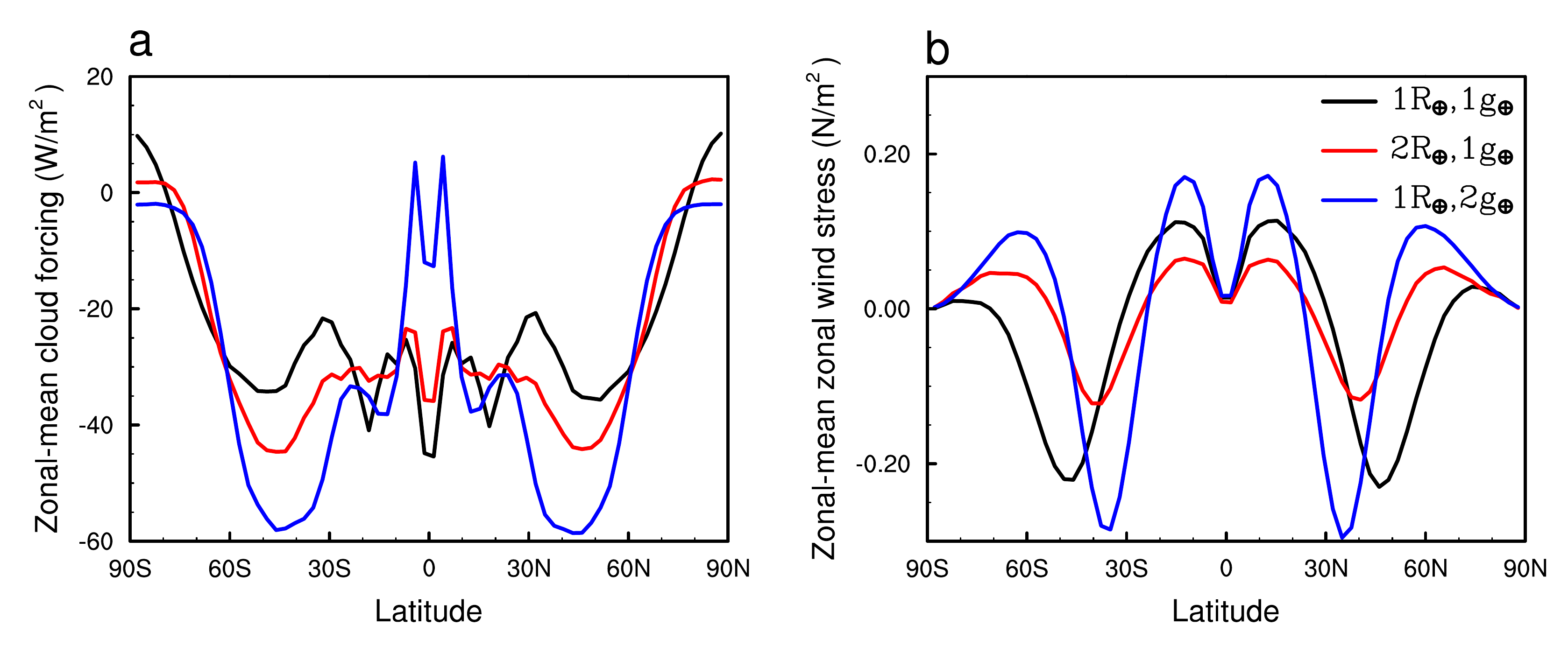}
\caption{(a) Zonal-mean cloud radiative effect (longwave plus shortwave) at the top of the atmosphere  and (b) zonal-mean zonal surface wind stresses, in the experiments without ice albedo feedback.}
\label{IceAlb}
\end{center}
\end{figure}






 \section{Conclusion} \label{sec:Con}
By employing one global climate model to examine the effects of planetary radius and gravity on the surface climate of rapidly rotating Earth-like aqua-planets, we are led to the following conclusions: 

\begin{itemize} 
\item For one planet with a larger radius, the meridional temperature gradient ($\frac{\partial T}{ R \partial \phi }$) becomes weaker mainly due to the geometrical effect. As a result, the atmospheric jets, Hadley cells, Ferrel cells, and mid-latitude eddies weaken, and the meridional energy transport decreases, which has a warming effect in the tropics but a cooling effect in high latitudes.

\item For one planet with a greater gravity, the surface temperature decreases everywhere mainly because water vapor content in the air decreases. The vertically-integrated water vapor mass is approximately inversely proportional to gravity. This conclusion could be applicable to other condensible greenhouse gases, such as CH$_4$ on Titan-like satellites, CO$_2$ on Mars-like planets, and N$_2$ on Pluto-like satellites.

\item When the radius and gravity are increased simultaneously, the surface temperature decreases nearly everywhere. This is due to  that the effect of varying gravity is stronger than that of varying radius in the simulations. Ice albedo feedback and water vapor feedback act to further amplify the response of the surface temperature. 
\end{itemize}  


These results suggest that future studies on the climate and habitability of exoplanets and on the inner and outer edges of the habitable zone should consider the effects of different radii and gravities.

\acknowledgments \textbf{Acknowledgments:} 
J.Y. acknowledges support from the National Science Foundation 
of China (NSFC) grants 41675071 and 41861124002.


\end{document}